\documentclass[aps,preprint,prd,showpacs,nofootinbib]{revtex4}
\usepackage{comment}
\usepackage{graphicx}
\usepackage{dcolumn}
\usepackage{bm}
\usepackage{hyperref}
\usepackage{graphicx} 
\usepackage{subfigure}
\usepackage{bm}
\usepackage{amsmath,amssymb}
\usepackage{latexsym}
\usepackage{ulem}
\usepackage{datetime}
\usepackage{scrtime}
\usepackage[dvipsnames, svgnames, x11names]{xcolor}

\definecolor{darkgreen}{RGB}{50,150,0}
\definecolor{purple}{cmyk}{0.5,0.75,0,0}
\definecolor{darkpurple}{RGB}{128,0,128}
\definecolor{ultramarine}{rgb}{0.07, 0.04, 0.56}
\definecolor{cadmiumgreen}{rgb}{0.0, 0.42, 0.24}
\definecolor{indigo(dye)}{rgb}{0.0, 0.25, 0.42}

\hypersetup{colorlinks=true,linkcolor=blue,citecolor=blue,urlcolor=blue}

\def\be{\begin{equation}}
\def\ee{\end{equation}}
\def\ba{\begin{eqnarray}}
\def\ea{\end{eqnarray}}

\def\lf{\left}
\def\rt{\right}
\def\d{\mathrm{d}}

\renewcommand{\[}{\left[}
\renewcommand{\]}{\right]}

\begin{document}
\title{Gravitational Waves from Primordial Black Holes formed by Null Energy Condition Violation during Inflation}

\author{Dong-Hui Yu$^{1}$}
\author{Jia-Zuo Zhang$^{1}$}
\author{Yong Cai$^{1}$}
\email[Corresponding author:~]{caiyong@zzu.edu.cn}
\affiliation{$^1$ Institute for Astrophysics, School of Physics, Zhengzhou University, Zhengzhou 450001, China}

\begin{abstract}
A transient violation of the null energy condition (NEC) during inflation provides a novel mechanism for producing primordial black holes (PBHs) and stochastic gravitational wave (GW) backgrounds. In this work, we extend previous studies by computing the GW contributions from both the ringdown phase of PBH formation and subsequent binary mergers. Our results show that this scenario produces a rich, multi-component GW spectrum consisting of primordial GWs, scalar-induced GWs, and GW emissions from PBH ringdown and binary mergers. We demonstrate that these correlated signatures across different frequency bands provide a novel and powerful avenue to probe or constrain NEC violation during inflation through future multi-band GW observations.
\end{abstract}

\maketitle

\section{Introduction}
\label{sec:intro}

Gravitational waves (GWs) serve as powerful probes of cosmic evolution and high-energy astrophysical processes. Among the most intriguing targets are primordial black holes (PBHs) \cite{Hawking:1971ei,Carr:1974nx,Carr:1975qj}, a unique class of compact objects formed in the early universe, which have been extensively discussed as potential dark matter candidates \cite{Ivanov:1994pa,Carr:2009jm,Belotsky:2014kca,Carr:2016drx,Carr:2020gox,Carr:2020xqk}. A widely studied formation mechanism is based on the inflationary paradigm with temporary deviations from slow-roll dynamics, most notably ultra-slow-roll phases during which the parameter $\epsilon \equiv -\dot{H}/H^2$ becomes sufficiently small to enhance curvature perturbations to amplitudes that can trigger gravitational collapse after horizon reentry \cite{Garcia-Bellido:2017mdw,Germani:2017bcs,Byrnes:2018txb,Fu:2019ttf,Fu:2019vqc,Fu:2020lob,Ragavendra:2020sop,Di:2017ndc,Motohashi:2017kbs,Dalianis:2018frf,Yi:2020cut}, see also e.g. \cite{Huang:2023chx,Choudhury:2023hfm,Choudhury:2024one,Chen:2024gqn,Wang:2024vfv,Choudhury:2024dei,Heydari:2024bxj,Wang:2024nmd,Calza:2024fzo,Calza:2024xdh,Calza:2025mwn,Domenech:2024rks,Domenech:2024cjn,Chen:2024pge,Iovino:2025cdy,Zhong:2025xwm,Ning:2026nfs,Ye:2026ffe,Gangopadhyay:2023qjr,Yogesh:2025hll,Mohammadi:2025avz,Allegrini:2024ooy,Allegrini:2025jha,Ashoorioon:2018uey,Ashoorioon:2019xqc,Ashoorioon:2020hln,Ashoorioon:2022raz}.

Recently, an interesting PBH formation mechanism has been proposed based on a transient violation of the null energy condition (NEC) during single-field inflation \cite{Cai:2023uhc}. In this scenario, the temporary NEC violation (with $\dot{H}>0$) drives a rapid growth of the Hubble parameter $H$ \cite{Cai:2020qpu}, which amplifies the power spectrum of primordial curvature perturbations on a certain range of scales and triggers gravitational collapse into PBHs upon horizon reentry. The enhanced curvature perturbations also inevitably generate a stochastic background of scalar-induced GWs (SIGWs). Moreover, the transient growth of $H$ directly amplifies the amplitude of primordial tensor perturbations, namely, primordial GWs (PGWs), on the corresponding scales \cite{Cai:2020qpu,Cai:2022nqv,Cai:2022lec,Jiang:2023gfe,Ye:2023tpz,Jiang:2024woi,Pan:2024ydt,Chen:2024mwg,Chen:2024jca,DOnofrio:2025bol}.

Furthermore, GW signals are expected to arise from both the ringdown phase of PBH formation and subsequent binary mergers \cite{DeLuca:2025uov,Yuan:2025bdp}, see also \cite{Phinney:2001di,Berti:2005ys,Bruegmann:2006ulg,Ajith:2007kx,Reisswig:2010di,Zhu:2011bd,Zhu:2012xw,Berti:2014bla,Raidal:2018bbj,Vaskonen:2019jpv,Cai:2019cdl,Hutsi:2020sol,Franciolini:2022tfm,Zhong:2024ysg,Raidal:2024bmm,Hu:2026sxj}. The relaxation of a newly formed PBH toward a stationary Kerr or Schwarzschild configuration corresponds to the ringdown phase, during which gravitational radiation is emitted in the form of quasinormal modes (QNMs) \cite{Berti:2009kk,Berti:2025hly}. In addition, PBHs may form binary systems during cosmic evolution, leading to further GW emission through binary coalescences.
Given the crucial roles of the NEC and its violations in cosmology (see e.g. \cite{Rubakov:2014jja,Curiel:2014zba,Kontou:2020bta} for reviews), these correlated signatures, namely SIGWs, PGWs, and GWs from PBH ringdown and binary mergers, collectively contribute to the stochastic GW background and provide a unique avenue to probe or constrain NEC violation in the primordial universe through multi-band GW observations. However, previous studies have focused only on the energy spectra of PGWs and SIGWs \cite{Cai:2023uhc}.

In this work, we extend previous studies of the scenario in which the primordial perturbations are enhanced by a transient NEC violation during inflation, producing PBHs, SIGWs and PGWs. We compute the resulting contributions to the stochastic GW background from both the ringdown phase of PBHs and subsequent binary mergers, including mergers arising from both two-body and three-body dynamical channels. We also assess the detectability of these various GW signals with current and future observatories.

This paper is organized as follows. In Sec.~\ref{sec:mechanism}, we briefly review the scenario of PBH formation induced by NEC violation during inflation. In Secs.~\ref{sec:ringdown} and \ref{sec:binary}, we present the formalism for computing the QNM signals from the ringdown phase and the GW energy spectra from binary PBH mergers, respectively. Sec.~\ref{sec:constraints} presents the resulting energy spectra for the various components of the GW background in our scenario and discusses their detectability. Finally, we summarize our results in Sec.~\ref{sec:conclusion}.

\section{PBHs and GWs from NEC Violation during Inflation}
\label{sec:mechanism}

In the context of nonsingular cosmology, it has been shown that a fully stable NEC violation can be realized within the framework of ``Beyond Horndeski'' theories \cite{Cai:2016thi,Creminelli:2016zwa,Cai:2017tku,Cai:2017dyi,Kolevatov:2017voe,Ilyas:2020qja}, see also e.g. \cite{Ijjas:2016tpn,Ijjas:2016vtq,Cai:2017dxl,Cai:2017pga,Cai:2022ori,Yu:2025wak,Mironov:2018oec,Ye:2019sth,Nandi:2019xag,Nandi:2020szp,Nandi:2023ooo,Zhu:2021whu,Zhu:2021ggm,Qiu:2024sdd,Ageeva:2024knc,Akama:2025ows,Dehghani:2025udv} for related studies. Following Ref. \cite{Cai:2023uhc}, we adopt the effective field theory (EFT) action
\be
\label{action-230112-1}
S=\int \d^4x\sqrt{-g}\Big[\frac{M_{\rm P}^2}{2} {\sf R} + P(\phi,X) + L_{\delta g^{00} {\sf R}^{(3)}} \Big] \,,
\ee
where $X \equiv \nabla_\mu\phi\nabla^\mu\phi$, the operator $L_{\delta g^{00} {\sf R}^{(3)}}=\frac{f(\phi)}{2}\delta g^{00} {\sf R}^{(3)}$ does not affect the background dynamics and is introduced to eliminate instabilities in the primordial perturbations \cite{Cai:2016thi,Cai:2017tku,Cai:2017dyi}. Here $\delta g^{00}$ denotes the perturbation of the $00$ component of the metric, and ${\sf R}^{(3)}$ is the three-dimensional Ricci scalar on the spacelike hypersurface.

The background evolution of our scenario is set by
\be
\label{260215P-1}
P(\phi,X) = - {g_1(\phi)\over 2} M_{\rm P}^2 X + {g_2(\phi)\over 4}X^2 -M_{\rm P}^4 V(\phi)\,,
\ee
where
\ba
	g_1(\phi)  &=&
	-{f_1 e^{2\phi} \over 1+f_1
		e^{2\phi} }+{2\over 1+e^{-q(\phi-\phi_0)}}+{1\over 1+e^{q(\phi-\phi_3)}} ~,
\\
	g_2(\phi) &=& {f_2\over 1+e^{-q(\phi-\phi_3)}} {1\over
		1+e^{q(\phi-\phi_0)}} ~,
\\
	V(\phi)  &=& \Lambda^4 \tanh ^{2}\left({\phi \over \sqrt{6 \alpha } }\right) {1\over
		1+e^{q(\phi-\phi_2)}}
	 +  \lambda \left[1-\frac{(\phi-\phi_1)^{2}}{\sigma^{2}}\right]^{2}{1\over
		1+e^{-p(\phi-\phi_1)}} ~.
\ea
The model is characterized by a set of positive constants $\{\lambda, \Lambda, \alpha, \sigma, f_{1,2}, p, q\}$. Furthermore, we consider a sequence of constant field values satisfying $\phi_3 < \phi_2 < 0 < \phi_1 < \phi_0$. The background evolution can be obtained by numerically solving the equations
\ba
3 M_{\rm P}^2 H^2&=& -2 \dot{\phi}^2 P_X-P\,,\\
M_{\rm P}^2 \dot{H} &=& \dot{\phi}^2 P_X .
\ea

With a suitable choice of parameters, we can realize a cosmological evolution where the universe transitions from an initial slow-roll inflationary phase at a lower energy scale $H \approx H_{\text{inf}1}$ to a subsequent one at a significantly higher scale $H \approx H_{\text{inf}2}\gg H_{\text{inf}1}$, bridged by an intermediate epoch of NEC violation \cite{Cai:2020qpu}.
In \cite{Cai:2023uhc}, four different sets of values were adopted for these model parameters. In this work, we continue to use the same four parameter sets.

To obtain the quadratic action of Eq. (\ref{action-230112-1}) for the primordial perturbations, we can use the $3+1$ decomposed metric $\d s^2=-N^2 \d t^2+h_{i j}\left(\d x^i+N^i \d t\right)\left(\d x^j+N^j \d t\right)$,
where $N$ is the lapse function, $N^i$ is the shift vector. In the unitary gauge, the 3-dimensional metric is $h_{ij}=a^2e^{2\zeta}(e^{\gamma})_{ij}$, in which $\gamma_{ii}=0=\partial_i\gamma_{ij}$.
Through a standard derivation, the quadratic action for tensor perturbation mode $\gamma_{ij}$ (i.e., PGWs) is given by
\begin{equation}
S_\gamma^{(2)}=\frac{M_{\rm P}^2}{8} \int \d^4 x a^3 \left[\dot{\gamma}_{i j}^2- \frac{\left(\partial_k \gamma_{i j}\right)^2}{a^2}\right]\,,
\end{equation}
which is same as that in general relativity.

In the momentum space,
\be \label{gamma0218}
\gamma_{ij}(\tau,\mathbf{x})=\int \frac{d^3k}{(2\pi)^{3}
}e^{-i\mathbf{k}\cdot \mathbf{x}} \sum_{\lambda=+,\times}
\hat{\gamma}_{\lambda}(\tau,\mathbf{k})
\epsilon^{(\lambda)}_{ij}(\mathbf{k}), \ee
in which
$\hat{\gamma}_{\lambda}(\tau,\mathbf{k})=
\gamma_{\lambda}(\tau,k)a_{\lambda}(\mathbf{k})
+\gamma_{\lambda}^*(\tau,-k)a_{\lambda}^{\dag}(-\mathbf{k})$;
the polarizations
$\epsilon_{ij}^{(\lambda)}(\mathbf{k})$ satisfy
$k_{j}\epsilon_{ij}^{(\lambda)}(\mathbf{k})=0$,
$\epsilon_{ii}^{(\lambda)}(\mathbf{k})=0$,
$\epsilon_{ij}^{(\lambda)}(\mathbf{k})
\epsilon_{ij}^{*(\lambda^{\prime}) }(\mathbf{k})=\delta_{\lambda
	\lambda^{\prime} }$ and $\epsilon_{ij}^{*(\lambda)
}(\mathbf{k})=\epsilon_{ij}^{(\lambda) }(-\mathbf{k})$;
$a_{\lambda}(\mathbf{k})$ and
$a^{\dag}_{\lambda}(\mathbf{k}^{\prime})$ satisfy $[
a_{\lambda}(\mathbf{k}),a_{\lambda^{\prime}}^{\dag}(\mathbf{k}^{\prime})
]=\delta_{\lambda\lambda^{\prime}}\delta^{(3)}(\mathbf{k}-\mathbf{k}^{\prime})$.
The power spectrum of PGWs is defined as
\be P_T\equiv {k^3\over 2\pi^2}\sum_{\lambda=+,\times}|\gamma_\lambda|^2\,.
\ee
The NEC violation during inflation leads to an enhancement of the PGW power spectrum at certain scales \cite{Cai:2020qpu,Cai:2022nqv,Ye:2023tpz,Pan:2024ydt}. In the presence of a parity-violating coupling to the background scalar field $\phi$, the NEC-violating phase significantly enhances both the parity-violating effect and the potential for detecting these effects within the PGW spectrum \cite{Cai:2022lec,Jiang:2024woi} (see also \cite{Cai:2016ihp,Zhu:2023lhv}).

According to \cite{Cai:2016thi}, the quadratic action of curvature perturbation mode $\zeta$ can be given as
\be
S_{\zeta}^{(2)}=\int \d^4x a^3 Q_s\lf[\dot{\zeta}^2-c_s^2
{(\partial\zeta)^2\over a^2} \rt] \,,\label{scalar-action-260215}
\ee
where
\be Q_s={2{\dot \phi}^4P_{XX}-M_{\rm P}^2{\dot H}\over H^2},\quad
c_s^2={M_{\rm P}^2 \over Q_s}\lf({{\dot c}_3\over a} -1\rt)\label{cs2}\,,
\ee
and $c_3=a(1+{2f\over M_{\rm P}^2})/H$.
Apparently, the ghost instability ($Q_s < 0$) can be avoided by a proper construction of $P(\phi, X)$, while the gradient instability (i.e., $c_s^2<0$) can be eliminated by the EFT operator $L_{\delta g^{00} {\sf R}^{(3)}}$.

In \cite{Cai:2023uhc}, it is demonstrated that a transient NEC violation during inflation can significantly amplify the primordial curvature power spectrum
\be P_{\zeta} \equiv \frac{k^3}{2\pi^2}|\zeta|^2\,.
\ee
Upon horizon reentry during the radiation-dominated era, these enhanced density perturbations undergo gravitational collapse to form PBHs with masses and abundances of observational interest, while inevitably generating a stochastic background of SIGWs, see \cite{Cai:2023uhc} for details.

\begin{figure}[ht]
    \centering
    \includegraphics[width=1\linewidth]{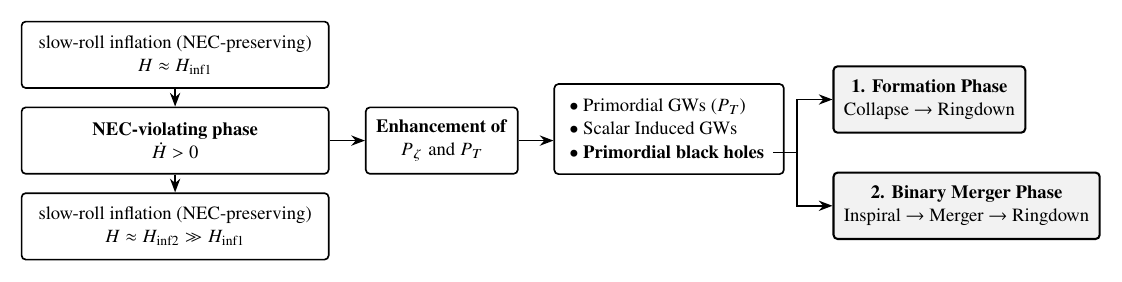}
    \caption{Schematic illustration of the GW generation framework driven by an intermediate NEC violation during inflation. The inflationary NEC-violating stage is characterized by an increasing Hubble parameter (i.e., $\dot{H}>0$), which triggers an enhancement of the primordial power spectra $P_\zeta$ and $P_T$.
This mechanism leads to three key predictions: PGWs, SIGWs and PBHs. PBHs further contribute to the GW background through two key channels: the initial ringdown during their formation and later-stage binary mergers.}
    \label{fig:mechanism}
\end{figure}

Intriguingly, in addition to the aforementioned signals, GW signals are also expected to arise from the ringdown phase of PBH formation and subsequent binary mergers \cite{DeLuca:2025uov,Yuan:2025bdp}.
To provide a comprehensive overview of our scenario and the resulting GW signatures, we illustrate the framework in Fig.~\ref{fig:mechanism}.

While previous studies \cite{Cai:2020qpu,Cai:2022nqv,Cai:2022lec,Cai:2023uhc} have addressed the PGWs and SIGWs arising from this scenario, this work focuses on the GW emissions associated with the PBH evolution itself. Specifically, we investigate the QNMs during the ringdown phase of PBH formation, as well as the signals from binary black hole mergers, considering both two-body and three-body formation channels.

\section{GWs from PBH ringdown}
\label{sec:ringdown}

The energy density spectrum of the PBH ringdown component within the stochastic GW background can be characterized by the two GW polarization states, i.e., $\gamma_+$ and $\gamma_\times$, which are defined below Eq. (\ref{gamma0218}). Consequently, the energy spectrum in the source frame can be expressed as \cite{Phinney:2001di,Zhu:2012xw}\footnote{In the following, we adopt natural units with $c = G_N = 1$.}
\begin{equation}
    \frac{\d E}{\d f_s}=\int \d \Omega \frac{\pi}{2}r^2 f_s^2\left(|\tilde{\gamma}_{+}(f_s,\theta,\phi)|^2+|\tilde{\gamma}_{\times}(f_s,\theta,\phi)|^2\right) \,,
    \label{eq:dEdfsringdown}
\end{equation}
where $\tilde{\gamma}_+$ and $\tilde{\gamma}_\times$ denote the Fourier transforms of the two polarization states, respectively. The relationship between the two polarization states and the Weyl scalar $\Psi_4$ is given by \cite{Berti:2005ys,Yuan:2025bdp}
\begin{equation}
    \Psi_4=\frac{1}{2}\left( \frac{\partial^2 \gamma_{+}}{\partial t^2}-\mathrm{i}\frac{\partial^2 \gamma_{\times}}{\partial t^2}\right) \,.
\end{equation}
To obtain the polarization states, it is necessary to integrate $\Psi_4$ twice with respect to time. To enforce causality and ensure that the GWs generated during the ringdown phase propagate along the light cone, we employ the Heaviside theta function \cite{Zhong:2024ysg} to truncate the Weyl scalar $\Psi_4$.

For gravitational perturbations, we assign a spin weight of $s=-2$. Solving the Teukolsky equation yields the Teukolsky master variable $\psi_{-2}$, and each QNM can be decomposed into radial and angular components as \cite{Berti:2014bla,Berti:2025hly,Berti:2009kk}
\begin{equation}
    \psi_{-2}(t,r,\theta,\phi)=S(\theta,\phi)R(r)e^{\mathrm{i} \omega t}~\,,
\end{equation}
where $\omega$ denotes the complex QNM frequency. In this work, we focus on non-spinning black holes and restrict our attention to the fundamental mode $M\omega = 0.3737-0.08896\mathrm{i}$ with $l=m=2$ \cite{Berti:2005ys,Yuan:2025bdp}. In this case, $S(\theta,\phi)$ reduces to the spin-weighted spherical harmonic with spin weight $s=-2$, given by ${}_{-2}Y_{22}(\theta,\phi)=\sqrt{\frac{5}{64\pi}}(1+\cos\theta)^2e^{2\mathrm{i}\phi}$ \cite{Bruegmann:2006ulg,Berti:2009kk,Berti:2025hly,Goldberg:1966uu}. The relationship between $\psi$ and the Weyl scalar is $\Psi_4=\psi_{-2}(t,r,\theta,\phi)r^{-4}$ \cite{Berti:2025hly,Berti:2009kk}.

In the following, we focus on monochromatic PBHs. In this case, the contribution of the ringdown phase to the power spectrum of stochastic GW background can be expressed as \cite{Yuan:2025bdp}
\begin{equation}
    h^2\Omega_{GW}(f)=h^2f\frac{f_\text{PBH}\Omega_\text{cdm}}{M}\left.\frac{\d E}{\d f_s}\right|_{f_s=f(1+z_{\text{form}})}~\,,
\end{equation}
where $\d E/\d f_s$ is the energy spectrum given by Eq. \eqref{eq:dEdfsringdown}, and $z_{\text{form}}$ denotes the redshift at the epoch of PBH formation. Here, we adopt the dimensionless Hubble parameter $h \simeq 0.674$, defined via $H_0 \equiv 100 h \, \mathrm{km \, s^{-1} \, Mpc^{-1}}$, with $H_0$ being the current Hubble constant; $f_{\text{PBH}} \equiv \Omega_{\text{PBH}}/\Omega_{\text{CDM}}$ denotes the PBH abundance; $\Omega_{\text{CDM}} = \rho_{\text{CDM}}/\rho_c$ is the dark matter density parameter, where $\rho_c = 3H_0^2 / (8\pi)$ represents the current critical energy density. The formation time $t_{\text{form}}$ is related to the PBH mass $M$ by \cite{Carr:1974nx,Carr:1975qj}
\begin{equation}
    M\simeq 2\times 10^{5}M_{\odot}\left(\frac{t_{\text{form}}}{1s}\right).
\end{equation}
The relationship between cosmic time and redshift is given by
\begin{equation}
    t=\int_z^{\infty}\frac{\d z}{H_0\sqrt{\[\Omega_r(1+z)^4+\Omega_m(1+z)^3+\Omega_{\Lambda}\]}(1+z)}~\,,
    \label{eq:ztotime}
\end{equation}
where $\Omega_r$, $\Omega_m$, and $\Omega_{\Lambda}$ represent the current density parameters for radiation, matter, and dark energy, respectively \cite{Planck:2018vyg}.

To quantitatively characterize the ringdown component of the stochastic GW background, we introduce a radiation efficiency parameter $\varepsilon$, which represents the fraction of the PBH mass converted into GWs during the ringdown phase \cite{Berti:2005ys,Yuan:2025bdp}. In our analysis, we adopt $\varepsilon=3\%$ and $\varepsilon=0.1\%$. The value $\varepsilon=3\%$ is chosen as an optimistic estimate, whereas $\varepsilon=0.1\%$ serves as a conservative estimate representing the lower bound for detectability \cite{Berti:2005ys}. The corresponding energy spectrum is required to satisfy the normalization condition
\begin{equation}
    \int_0^{+\infty}\frac{\d E}{\d f_s}\d f_s=\varepsilon M.
\end{equation}

In the computation of the energy spectrum $\d E/\d f_s$, we observe non-physical numerical artifacts in the low-frequency regime ($Mf < 0.04$), which arise from the inherent limitations of the numerical Fourier transform. These artifacts manifest specifically as an anomalous divergence in the spectrum. Causality constraints dictate that for a temporally localized signal, the GW energy density spectrum must follow the asymptotic scaling law $\Omega_{GW} \propto f^3$ in the infrared limit \cite{Cai:2019cdl}. Consequently, the single-source energy spectrum is required to satisfy $\d E/\d f_s \propto f^2$. To mitigate these low-frequency numerical errors, Ref. \cite{Reisswig:2010di} proposed a method involving a hard cutoff between the physical and non-physical regimes, followed by fixed-frequency integration. However, such a hard truncation approach fails to accurately reproduce the theoretically expected asymptotic behavior in the limit $f \to 0$.

To address this issue, we employ a hybrid matching scheme. In the low-frequency regime (defined by a cutoff at $Mf_{\text{cut}} < 0.05$), we replace the numerical results with an analytical solution derived from the approximation of slowly-damped modes
\begin{equation}
\begin{split}
    \frac{\d E}{\d f_s} &= \frac{Af_s^2}{\tau^2} \bigg\{ \left[(f_s+f_0)^2+(2\pi \tau)^{-2}\right]^2 + \left[(f_s-f_0)^2+(2\pi \tau)^{-2}\right]^2 \bigg\}^{-1} \,,
\end{split}
\label{eq:energy_spectrum_fixed}
\end{equation}
where $A$ is a constant dependent on the PBH mass $M$ \cite{Berti:2005ys}. For the $l=m=2$ QNM, the oscillation frequency $f_0$ and the damping time $\tau$ are given by $f_0 = \frac{\text{Re}[\omega]}{2\pi}$ and $\tau = \left| \frac{1}{\text{Im}[\omega]} \right|$, respectively.
This analytical tail is then stitched to the numerical results using a smooth transition function. Crucially, this analytical form strictly satisfies the scaling relation $\d E/\d f_s \propto f^2$ in the limit $f \to 0$, thereby ensuring the physical self-consistency of the energy spectrum across the entire frequency band.

\section{GWs from binary PBH coalescences}
\label{sec:binary}

The energy spectrum of the stochastic GW background arising from binary black hole mergers can be expressed as
\begin{equation}
\begin{split}
    h^2\Omega_\text{GW}(f) &= \frac{fh^2}{\rho_c} \int \d m_1 \, \d m_2 \, \d z \, \frac{\d t}{\d z}\times R(z,m_1,m_2,m_3) \left.\frac{\d E}{\d f_s}\right|_{f_s=f(1+z)} \,,
\end{split}
\label{eq:omega_gw}
\end{equation}
where $m_1$ and $m_2$ denote the masses of the merging binary components, while $m_3$ represents the mass of the third PBH ejected during the three-body interaction \cite{Raidal:2024bmm}. The derivative ${\d t}/{\d z}$ is determined using the time-redshift relation given by Eq. \eqref{eq:ztotime}. The quantity $R(z, m_1, m_2, m_3)$ denotes the total merger rate density.

While the total merger rate is generally composed of four distinct contributions as detailed in Ref. \cite{Raidal:2024bmm}, we focus primarily on two channels in this work: the early two-body channel and the early three-body channel, see also \cite{Yuan:2025bdp}. Their corresponding merger rate densities are denoted by $R_2(z, m_1, m_2)$ and $R_3(z, m_1, m_2,m_3)$, respectively. Consequently, the total merger rate density is given by the sum, i.e.,
\begin{equation} R(z,m_1,m_2,m_3)=R_2(z,m_1,m_2)+R_3(z,m_1,m_2,m_3).
\end{equation}

The merger rate for the two-body channel, i.e., $R_2(z, m_1,m_2)$, is expressed as \cite{Franciolini:2022tfm,Hutsi:2020sol}
\begin{equation}
\begin{split}
    R_2(z,m_1,m_2) &= \frac{1.6\times 10^6}{\mathrm{Gpc^3\,yr}} f_{\text{PBH}}^{\frac{53}{37}} \left(\frac{t(z)}{t_0}\right)^{-\frac{34}{37}} \times \eta^{-\frac{34}{37}} \left(\frac{M}{M_{\odot}}\right)^{-\frac{32}{37}} \times S[M,f_{\text{PBH}},\psi,z] \psi(m_1)\psi(m_2) \,,
\end{split}
\label{eq:R2_rate}
\end{equation}
where $M = m_1 + m_2$ denotes the total mass of the binary system, $\eta = m_1 m_2 / M^2$ is the symmetric mass ratio, and $t_0$ represents the current age of the universe, as determined by Eq. (\ref{eq:ztotime}). Furthermore, $\psi$ denotes the PBH mass distribution, normalized via $\int \psi(m)\d m=1$, and the term $S[M, f_{\text{PBH}}, \psi, z]$ serves as a suppression factor (see Ref. \cite{Hutsi:2020sol} for details).

The merger rate for the three-body channel, i.e., $R_3(z,m_1,m_2,m_3)$, is expressed as \cite{Raidal:2024bmm}
\begin{equation}
\begin{split}
    R_3(z,m_1,m_2,m_3) &= \frac{7.9\times10^4}{\mathrm{Gpc^3\,yr}} \left(\frac{t}{t_0}\right)^{\frac{\gamma}{7}-1} f_{\text{PBH}}^{\frac{144\gamma}{259}+\frac{47}{37}}  \times \left[\frac{\langle m \rangle}{M_{\odot}}\right]^{\frac{5\gamma-32}{37}} \left(\frac{M}{2\langle m \rangle }\right)^{\frac{179\gamma}{259}-\frac{2122}{333}} \\
    &\times (4\eta)^{-\frac{3\gamma}{7}-1} \mathcal{K} \frac{e^{-3.2(\gamma-1)}\gamma}{28/9-\gamma}  \bar{\mathcal{F}}(m_1,m_2)\psi(m_1)\psi(m_2) \,,
\end{split}
\label{eq:R3_optimized}
\end{equation}
where $\gamma$ and $\mathcal{K}$ quantify the angular momentum distribution and the binary hardening effect due to binary-single collisions, respectively. For the numerical analysis \cite{Raidal:2024bmm,Raidal:2018bbj}, we set $\gamma=1$ and $\mathcal{K}=4$. The mean mass of the PBH distribution can be expressed as $\left \langle m \right \rangle=\int m\psi(m)\d m$. The $\bar{\mathcal{F}}$ function can be defined as
\begin{equation}
\begin{split}
    \bar{\mathcal{F}}(m_1,m_2) &\equiv \int \d m \, \psi(m) \frac{\langle m \rangle}{m} \times \left[2\mathcal{F}(m_1,m_2,m)+\mathcal{F}(m,m_1,m_2)\right] \,,
\end{split}
\label{eq:F_bar_integral}
\end{equation}
with the definition of the dimensionless factor
\begin{equation}
\begin{split}
    \mathcal{F}(m_1,m_2,m_3) &\equiv m_1^{5/3}m_2^{5/3}m_3^{7/9}  \times \left(\frac{m_1+m_2}{2}\right)^{4/9} \left(\frac{m_1+m_2+m_3}{3}\right)^{2/9}  \times \langle m \rangle^{-43/9}.
\end{split}
\label{eq:F_function}
\end{equation}
For a monochromatic PBH population characterized by the mass function $\psi(m') = \delta(m' - m)$, which implies $m_1 = m_2 = m_3 = m$ \cite{Vaskonen:2019jpv}, the parameters defined above simplify significantly. Specifically, we obtain $\langle m \rangle = m$, $\mathcal{F}=1$ and $\bar{\mathcal{F}}=3$. Consequently, the total mass and symmetric mass ratio for the binary system reduce to $M = 2m$ and $\eta = 1/4$, respectively.

The coalescence process of binary PBHs consists of three phases: inspiral, merger, and ringdown, each characterized by a distinct energy spectrum. In the source frame, the spectrum can be expressed as \cite{Zhu:2011bd}
\begin{equation}
\begin{split}
    \frac{\d E}{\d f_s} &= \frac{(G\pi)^{2/3} M_c^{5/3}}{3}  \times \begin{cases}
        f_s^{-1/3}, & \text{if } f_s < f_1 \\
        \dfrac{f_s^{2/3}}{f_1}, & \text{if } f_1 \le f_s < f_2 \\
        \dfrac{f_s \, (f_1 f_2^{4/3})^{-1}}{1 + \left[\frac{2(f_s-f_2)}{\sigma}\right]^2}, & \text{if } f_2 \le f_s < f_3
    \end{cases}
\end{split}
\label{eq:single_col_cases_with_if}
\end{equation}
where $M_c$ denotes the chirp mass, which reduces to $M_c = m/2^{1/5}$ for monochromatic PBHs. These spectral parameters, denoted by $\{f_1, f_2, \sigma, f_3\}$, are determined by the binary properties $M$ and $\eta$ via the relation $(a\eta^2 + b\eta + c)/(\pi M)$, with the coefficients $(a, b, c)$ adopted from Ref. \cite{Ajith:2007kx}.

\section{Multi-component stochastic GW background spectra and observational constraints}
\label{sec:constraints}

\begin{figure}[htbp]
    \centering
\includegraphics[width=0.85\linewidth]{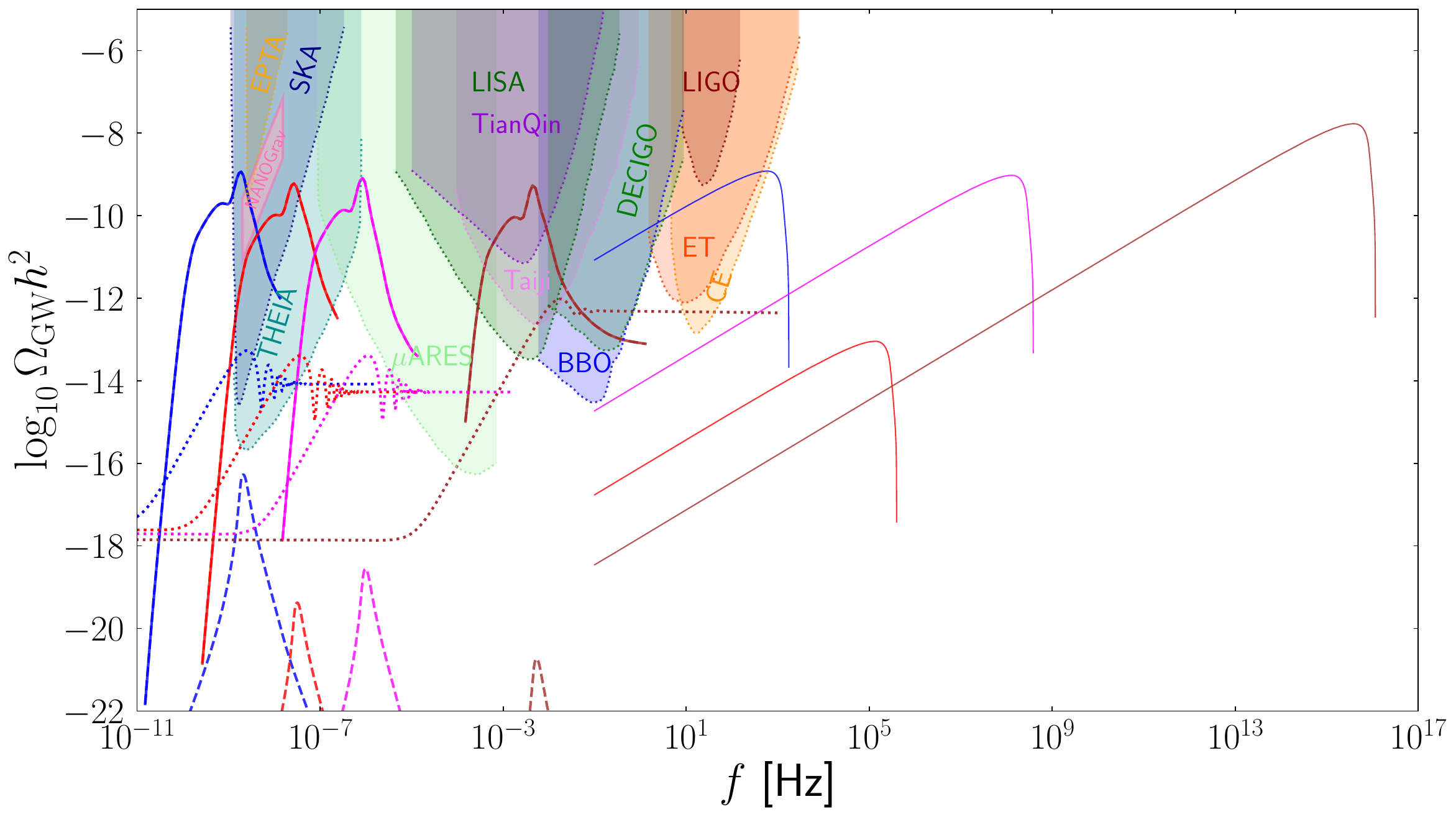}
\caption{Predicted energy spectra of the stochastic GW background for four different parameter sets (distinguished by color). These curves correspond to those of the same colors in Figs. 3 to 5 of Ref. \cite{Cai:2023uhc}, namely, curves of the same color share identical parameter settings.
The various line styles denote different generation mechanisms:
thick solid curves for SIGWs;
dotted curves for PGWs;
dashed curves for the QNM radiation during the ringdown phase of newly
formed PBHs;
and thin solid curves for the binary PBH mergers. Here, the results for SIGWs and PGWs are adopted from Fig. 5 of Ref. \cite{Cai:2023uhc}.
The shaded regions and boundary curves represent sensitivities of current and future GW observatories, including PTAs (EPTA, NANOGrav, SKA), space-based detectors (THEIA, $\mu$ARES, LISA, Taiji, TianQin, BBO, DECIGO), and ground-based interferometers (LIGO, ET, CE), see e.g. \cite{Roshan:2024qnv} and references therein.}
    \label{fig:GWSpectrum}
\end{figure}

Following the methods introduced in Secs. \ref{sec:ringdown} and \ref{sec:binary}, we numerically calculate the GW signals arising from both the PBH ringdown phase and subsequent binary mergers within the scenario described in Sec. \ref{sec:mechanism}. The results are presented in Fig.~\ref{fig:GWSpectrum}, in which the curves distinguished by four colors correspond to the four sets of parameter values used for the curves of the same colors in Figs. 3 to 5 of Ref. \cite{Cai:2023uhc}. For comparison, the spectra of SIGWs and PGWs are also plotted in Fig.~\ref{fig:GWSpectrum}, with the results adopted from Fig. 5 of Ref.~\cite{Cai:2023uhc}.
We interpret the spectral features in Fig.~\ref{fig:GWSpectrum} to highlight the critical implications of our scenario.

In the case of solar-mass PBHs ($10^0-10^1 M_\odot$), the benchmark scenario (blue curves) offers a compelling opportunity for multi-band observations. While the low-frequency SIGWs could fall within the observation window of Pulsar Timing Arrays (PTAs), our model also predicts a high-frequency counterpart arising from PBH binary mergers. This inspiral signal (exhibiting a $f^{2/3}$ scaling) falls within the sensitivity windows of next-generation detectors, including DECIGO, BBO, ET and CE. In principle, a joint detection of PGWs, SIGWs and the high-frequency signals from PBH mergers would provide powerful insights into our scenario and help distinguish it from other PBH formation mechanisms.

To assess the robustness of the predicted binary merger signals, we explicitly incorporate the contribution from PBH mergers arising in three-body formation channels. While the dominance of three-body dynamics generally depends on the PBH abundance ($f_{\text{PBH}}$) and mass function \cite{Raidal:2024bmm}, we find that within the specific parameter space adopted by this work, these channels introduce only minor corrections to the standard early two-body baseline. Consequently, the spectra presented in Fig.~\ref{fig:GWSpectrum} (which include the full three-body results) represent a robust prediction, demonstrating that our conclusions are largely insensitive to the complexities of multi-body formation dynamics within the parameter space considered in this work.

At ultra-high frequencies, the GW background generated by PBH binaries may be constrained by Big Bang Nucleosynthesis (BBN). For subsolar asteroid-mass PBHs ($M \sim 10^{-12} M_\odot$), the GW spectrum from PBH mergers peaks around $\sim 10^{15}$ Hz (brown curve), far beyond the reach of conventional interferometers. In this regime, BBN constrains the additional radiation energy density contributed by GWs present during nucleosynthesis. Our calculation yields an integrated energy density of $\int \Omega_{\rm GW} h^2 \, \d\ln f \sim 10^{-7}$, which remains well within the current $2\sigma$ bound imposed by $N_{\rm eff}$ measurements ($\lesssim 4.4 \times 10^{-6}$) \cite{Cooke:2013cba,Clarke:2020bil}. Considering that only the early portion of mergers occurs before or during the BBN epoch, this consistency ensures the viability of our model and highlights that precision cosmology can effectively constrain PBH physics.

However, despite the promising signals discussed above, certain GW components predicted in our scenario face significant observational challenges. For the magenta and red curves, which correspond to PBH masses of $M \sim 10^{-5} M_\odot$ and $10^{-2} M_\odot$ respectively, the peaks of the binary merger signals are located at $\sim 10^8$ Hz and $\sim 10^5$ Hz. Despite these high characteristic frequencies, their low-frequency spectral tails possess insufficient amplitude to reach the sensitivity thresholds of space-based detectors.  Furthermore, regarding the GWs emitted during the newly formed PBH ringdown phase, we find that their peak frequency coincides in order of magnitude with that of the SIGWs.
Consequently, this component is inevitably overshadowed by the dominant SIGW and PGW backgrounds. Notably, while our results assume an optimistic radiation efficiency of $\varepsilon=3\%$, a more conservative estimate (e.g., $\varepsilon=0.1\%$) would make the formation ringdown signal even more negligible.

\section{Conclusion}
\label{sec:conclusion}

In this work, we have investigated the multi-component stochastic GW background generated by a transient NEC violation during inflation. While previous studies primarily focused on the spectra of PGWs and SIGWs, we extended the analysis to include GW signals directly associated with the evolution of PBHs. Specifically, we calculated the gravitational radiation from the QNMs during the ringdown phase of newly formed PBHs and the signals from subsequent binary PBH mergers, accounting for both two-body and three-body formation channels.

Our results demonstrate that this scenario can produce a rich variety of GW signals across a wide frequency range. For solar-mass PBHs, our model predicts a compelling multi-band signature where low-frequency SIGWs and high-frequency binary merger signals could be detected by PTAs and next-generation ground-based interferometers, respectively. We found that the binary merger signals are robust against three-body dynamics, and the ultra-high frequency GWs from asteroid-mass PBHs remain consistent with BBN constraints. However, the analysis also suggests that the GWs from the initial PBH ringdown are typically overshadowed by the dominant SIGWs and PGWs, making them challenging to distinguish.

In summary, this study indicates that NEC violation during inflation can generate a unique combination of GW backgrounds, comprising PGWs, SIGWs, and PBH-related emissions from ringdown and mergers. These correlated signals provide a novel avenue to probe or constrain NEC violation during inflation. With the rapid development of multi-band GW astronomy, such distinct spectral features offer a promising strategy for future observations to verify the physics of the primordial universe.

\acknowledgments

This work is supported in part by the National Natural Science Foundation of China (Grant No. 12575066) and the Natural Science Foundation of Henan Province (Grant Nos. 262300421236, 242300420231). The data that support the findings of this article are openly available \cite{data260220}.

\bibliography{SGWB}

\providecommand{\href}[2]{#2}\begingroup\raggedright\begin{thebibliography}{100}

\bibitem{Hawking:1971ei}
S.~Hawking, ``{Gravitationally collapsed objects of very low mass},''
  \href{http://dx.doi.org/10.1093/mnras/152.1.75}{{\em Mon. Not. Roy. Astron.
  Soc.} {\bfseries 152} (1971) 75}.

\bibitem{Carr:1974nx}
B.~J. Carr and S.~W. Hawking, ``{Black holes in the early Universe},''
  \href{http://dx.doi.org/10.1093/mnras/168.2.399}{{\em Mon. Not. Roy. Astron.
  Soc.} {\bfseries 168} (1974) 399--415}.

\bibitem{Carr:1975qj}
B.~J. Carr, ``{The Primordial black hole mass spectrum},''
  \href{http://dx.doi.org/10.1086/153853}{{\em Astrophys. J.} {\bfseries 201}
  (1975) 1--19}.

\bibitem{Ivanov:1994pa}
P.~Ivanov, P.~Naselsky, and I.~Novikov, ``{Inflation and primordial black holes
  as dark matter},'' \href{http://dx.doi.org/10.1103/PhysRevD.50.7173}{{\em
  Phys. Rev. D} {\bfseries 50} (1994) 7173--7178}.

\bibitem{Carr:2009jm}
B.~J. Carr, K.~Kohri, Y.~Sendouda, and J.~Yokoyama, ``{New cosmological
  constraints on primordial black holes},''
  \href{http://dx.doi.org/10.1103/PhysRevD.81.104019}{{\em Phys. Rev. D}
  {\bfseries 81} (2010) 104019},
  \href{http://arxiv.org/abs/0912.5297}{{\ttfamily arXiv:0912.5297
  [astro-ph.CO]}}.

\bibitem{Belotsky:2014kca}
K.~M. Belotsky, A.~D. Dmitriev, E.~A. Esipova, V.~A. Gani, A.~V. Grobov, M.~Y.
  Khlopov, A.~A. Kirillov, S.~G. Rubin, and I.~V. Svadkovsky, ``{Signatures of
  primordial black hole dark matter},''
  \href{http://dx.doi.org/10.1142/S0217732314400057}{{\em Mod. Phys. Lett. A}
  {\bfseries 29} no.~37, (2014) 1440005},
  \href{http://arxiv.org/abs/1410.0203}{{\ttfamily arXiv:1410.0203
  [astro-ph.CO]}}.

\bibitem{Carr:2016drx}
B.~Carr, F.~Kuhnel, and M.~Sandstad, ``{Primordial Black Holes as Dark
  Matter},'' \href{http://dx.doi.org/10.1103/PhysRevD.94.083504}{{\em Phys.
  Rev. D} {\bfseries 94} no.~8, (2016) 083504},
  \href{http://arxiv.org/abs/1607.06077}{{\ttfamily arXiv:1607.06077
  [astro-ph.CO]}}.

\bibitem{Carr:2020gox}
B.~Carr, K.~Kohri, Y.~Sendouda, and J.~Yokoyama, ``{Constraints on primordial
  black holes},'' \href{http://dx.doi.org/10.1088/1361-6633/ac1e31}{{\em Rept.
  Prog. Phys.} {\bfseries 84} no.~11, (2021) 116902},
  \href{http://arxiv.org/abs/2002.12778}{{\ttfamily arXiv:2002.12778
  [astro-ph.CO]}}.

\bibitem{Carr:2020xqk}
B.~Carr and F.~Kuhnel, ``{Primordial Black Holes as Dark Matter: Recent
  Developments},''
  \href{http://dx.doi.org/10.1146/annurev-nucl-050520-125911}{{\em Ann. Rev.
  Nucl. Part. Sci.} {\bfseries 70} (2020) 355--394},
  \href{http://arxiv.org/abs/2006.02838}{{\ttfamily arXiv:2006.02838
  [astro-ph.CO]}}.

\bibitem{Garcia-Bellido:2017mdw}
J.~Garcia-Bellido and E.~Ruiz~Morales, ``{Primordial black holes from single
  field models of inflation},''
  \href{http://dx.doi.org/10.1016/j.dark.2017.09.007}{{\em Phys. Dark Univ.}
  {\bfseries 18} (2017) 47--54},
  \href{http://arxiv.org/abs/1702.03901}{{\ttfamily arXiv:1702.03901
  [astro-ph.CO]}}.

\bibitem{Germani:2017bcs}
C.~Germani and T.~Prokopec, ``{On primordial black holes from an inflection
  point},'' \href{http://dx.doi.org/10.1016/j.dark.2017.09.001}{{\em Phys. Dark
  Univ.} {\bfseries 18} (2017) 6--10},
  \href{http://arxiv.org/abs/1706.04226}{{\ttfamily arXiv:1706.04226
  [astro-ph.CO]}}.

\bibitem{Byrnes:2018txb}
C.~T. Byrnes, P.~S. Cole, and S.~P. Patil, ``{Steepest growth of the power
  spectrum and primordial black holes},''
  \href{http://dx.doi.org/10.1088/1475-7516/2019/06/028}{{\em JCAP} {\bfseries
  06} (2019) 028}, \href{http://arxiv.org/abs/1811.11158}{{\ttfamily
  arXiv:1811.11158 [astro-ph.CO]}}.

\bibitem{Fu:2019ttf}
C.~Fu, P.~Wu, and H.~Yu, ``{Primordial Black Holes from Inflation with
  Nonminimal Derivative Coupling},''
  \href{http://dx.doi.org/10.1103/PhysRevD.100.063532}{{\em Phys. Rev. D}
  {\bfseries 100} no.~6, (2019) 063532},
  \href{http://arxiv.org/abs/1907.05042}{{\ttfamily arXiv:1907.05042
  [astro-ph.CO]}}.

\bibitem{Fu:2019vqc}
C.~Fu, P.~Wu, and H.~Yu, ``{Scalar induced gravitational waves in inflation
  with gravitationally enhanced friction},''
  \href{http://dx.doi.org/10.1103/PhysRevD.101.023529}{{\em Phys. Rev. D}
  {\bfseries 101} no.~2, (2020) 023529},
  \href{http://arxiv.org/abs/1912.05927}{{\ttfamily arXiv:1912.05927
  [astro-ph.CO]}}.

\bibitem{Fu:2020lob}
C.~Fu, P.~Wu, and H.~Yu, ``{Primordial black holes and oscillating
  gravitational waves in slow-roll and slow-climb inflation with an
  intermediate noninflationary phase},''
  \href{http://dx.doi.org/10.1103/PhysRevD.102.043527}{{\em Phys. Rev. D}
  {\bfseries 102} no.~4, (2020) 043527},
  \href{http://arxiv.org/abs/2006.03768}{{\ttfamily arXiv:2006.03768
  [astro-ph.CO]}}.

\bibitem{Ragavendra:2020sop}
H.~V. Ragavendra, P.~Saha, L.~Sriramkumar, and J.~Silk, ``{Primordial black
  holes and secondary gravitational waves from ultraslow roll and punctuated
  inflation},'' \href{http://dx.doi.org/10.1103/PhysRevD.103.083510}{{\em Phys.
  Rev. D} {\bfseries 103} no.~8, (2021) 083510},
  \href{http://arxiv.org/abs/2008.12202}{{\ttfamily arXiv:2008.12202
  [astro-ph.CO]}}.

\bibitem{Di:2017ndc}
H.~Di and Y.~Gong, ``{Primordial black holes and second order gravitational
  waves from ultra-slow-roll inflation},''
  \href{http://dx.doi.org/10.1088/1475-7516/2018/07/007}{{\em JCAP} {\bfseries
  07} (2018) 007}, \href{http://arxiv.org/abs/1707.09578}{{\ttfamily
  arXiv:1707.09578 [astro-ph.CO]}}.

\bibitem{Motohashi:2017kbs}
H.~Motohashi and W.~Hu, ``{Primordial Black Holes and Slow-Roll Violation},''
  \href{http://dx.doi.org/10.1103/PhysRevD.96.063503}{{\em Phys. Rev. D}
  {\bfseries 96} no.~6, (2017) 063503},
  \href{http://arxiv.org/abs/1706.06784}{{\ttfamily arXiv:1706.06784
  [astro-ph.CO]}}.

\bibitem{Dalianis:2018frf}
I.~Dalianis, A.~Kehagias, and G.~Tringas, ``{Primordial black holes from
  {\ensuremath{\alpha}}-attractors},''
  \href{http://dx.doi.org/10.1088/1475-7516/2019/01/037}{{\em JCAP} {\bfseries
  01} (2019) 037}, \href{http://arxiv.org/abs/1805.09483}{{\ttfamily
  arXiv:1805.09483 [astro-ph.CO]}}.

\bibitem{Yi:2020cut}
Z.~Yi, Q.~Gao, Y.~Gong, and Z.-h. Zhu, ``{Primordial black holes and
  scalar-induced secondary gravitational waves from inflationary models with a
  noncanonical kinetic term},''
  \href{http://dx.doi.org/10.1103/PhysRevD.103.063534}{{\em Phys. Rev. D}
  {\bfseries 103} no.~6, (2021) 063534},
  \href{http://arxiv.org/abs/2011.10606}{{\ttfamily arXiv:2011.10606
  [astro-ph.CO]}}.

\bibitem{Huang:2023chx}
H.-L. Huang, Y.~Cai, J.-Q. Jiang, J.~Zhang, and Y.-S. Piao, ``{Supermassive
  Primordial Black Holes for Nano-Hertz Gravitational Waves and High-redshift
  JWST Galaxies},'' \href{http://dx.doi.org/10.1088/1674-4527/ad683d}{{\em Res.
  Astron. Astrophys.} {\bfseries 24} no.~9, (2024) 091001},
  \href{http://arxiv.org/abs/2306.17577}{{\ttfamily arXiv:2306.17577 [gr-qc]}}.

\bibitem{Choudhury:2023hfm}
S.~Choudhury, A.~Karde, S.~Panda, and M.~Sami, ``{Scalar induced gravity waves
  from ultra slow-roll galileon inflation},''
  \href{http://dx.doi.org/10.1016/j.nuclphysb.2024.116678}{{\em Nucl. Phys. B}
  {\bfseries 1007} (2024) 116678},
  \href{http://arxiv.org/abs/2308.09273}{{\ttfamily arXiv:2308.09273
  [astro-ph.CO]}}.

\bibitem{Choudhury:2024one}
S.~Choudhury, A.~Karde, S.~Panda, and M.~Sami, ``{Realisation of the ultra-slow
  roll phase in Galileon inflation and PBH overproduction},''
  \href{http://dx.doi.org/10.1088/1475-7516/2024/07/034}{{\em JCAP} {\bfseries
  07} (2024) 034}, \href{http://arxiv.org/abs/2401.10925}{{\ttfamily
  arXiv:2401.10925 [astro-ph.CO]}}.

\bibitem{Chen:2024gqn}
L.-Y. Chen, H.~Yu, and P.~Wu, ``{Resonant amplification of curvature
  perturbations in inflation model with periodical derivative coupling},''
  \href{http://dx.doi.org/10.1016/j.physletb.2024.138457}{{\em Phys. Lett. B}
  {\bfseries 849} (2024) 138457},
  \href{http://arxiv.org/abs/2401.07523}{{\ttfamily arXiv:2401.07523 [gr-qc]}}.

\bibitem{Wang:2024vfv}
X.~Wang, Y.-l. Zhang, and M.~Sasaki, ``{Enhanced curvature perturbation and
  primordial black hole formation in two-stage inflation with a break},''
  \href{http://dx.doi.org/10.1088/1475-7516/2024/07/076}{{\em JCAP} {\bfseries
  07} (2024) 076}, \href{http://arxiv.org/abs/2404.02492}{{\ttfamily
  arXiv:2404.02492 [astro-ph.CO]}}.

\bibitem{Choudhury:2024dei}
S.~Choudhury, A.~Karde, S.~Panda, and S.~SenGupta,
  ``{Regularized-renormalized-resummed loop corrected power spectrum of
  non-singular bounce with Primordial Black Hole formation},''
  \href{http://dx.doi.org/10.1140/epjc/s10052-024-13460-8}{{\em Eur. Phys. J.
  C} {\bfseries 84} no.~11, (2024) 1149},
  \href{http://arxiv.org/abs/2405.06882}{{\ttfamily arXiv:2405.06882
  [astro-ph.CO]}}.

\bibitem{Heydari:2024bxj}
S.~Heydari and K.~Karami, ``{Primordial Black Holes Generated by Fast-roll
  Mechanism in Noncanonical Natural Inflation},''
  \href{http://dx.doi.org/10.3847/1538-4357/ad7605}{{\em Astrophys. J.}
  {\bfseries 975} no.~1, (2024) 148},
  \href{http://arxiv.org/abs/2405.08563}{{\ttfamily arXiv:2405.08563 [gr-qc]}}.

\bibitem{Wang:2024nmd}
X.~Wang, X.-H. Ma, and Y.-F. Cai, ``{Primordial black hole formation from the
  upward step model: Avoiding overproduction},''
  \href{http://dx.doi.org/10.1142/S0218271825500270}{{\em Int. J. Mod. Phys. D}
  {\bfseries 34} no.~06, (2025) 2550027},
  \href{http://arxiv.org/abs/2412.19631}{{\ttfamily arXiv:2412.19631
  [astro-ph.CO]}}.

\bibitem{Calza:2024fzo}
M.~Calz{\`a}, D.~Pedrotti, and S.~Vagnozzi, ``{Primordial regular black holes
  as all the dark matter. I. Time-radial-symmetric metrics},''
  \href{http://dx.doi.org/10.1103/PhysRevD.111.024009}{{\em Phys. Rev. D}
  {\bfseries 111} no.~2, (2025) 024009},
  \href{http://arxiv.org/abs/2409.02804}{{\ttfamily arXiv:2409.02804 [gr-qc]}}.

\bibitem{Calza:2024xdh}
M.~Calz{\`a}, D.~Pedrotti, and S.~Vagnozzi, ``{Primordial regular black holes
  as all the dark matter. II. Non-time-radial-symmetric and loop quantum
  gravity-inspired metrics},''
  \href{http://dx.doi.org/10.1103/PhysRevD.111.024010}{{\em Phys. Rev. D}
  {\bfseries 111} no.~2, (2025) 024010},
  \href{http://arxiv.org/abs/2409.02807}{{\ttfamily arXiv:2409.02807 [gr-qc]}}.

\bibitem{Calza:2025mwn}
M.~Calz{\`a}, D.~Pedrotti, G.-W. Yuan, and S.~Vagnozzi, ``{Primordial regular
  black holes as all the dark matter. III. Covariant canonical quantum gravity
  models},'' \href{http://dx.doi.org/10.1103/4x1f-vctx}{{\em Phys. Rev. D}
  {\bfseries 112} no.~12, (2025) 124015},
  \href{http://arxiv.org/abs/2507.02396}{{\ttfamily arXiv:2507.02396 [gr-qc]}}.

\bibitem{Domenech:2024rks}
G.~Dom{\`e}nech, S.~Pi, A.~Wang, and J.~Wang, ``{Induced gravitational wave
  interpretation of PTA data: a complete study for general equation of
  state},'' \href{http://dx.doi.org/10.1088/1475-7516/2024/08/054}{{\em JCAP}
  {\bfseries 08} (2024) 054}, \href{http://arxiv.org/abs/2402.18965}{{\ttfamily
  arXiv:2402.18965 [astro-ph.CO]}}.

\bibitem{Domenech:2024cjn}
G.~Dom{\`e}nech and M.~Sasaki, ``{Probing primordial black hole scenarios with
  terrestrial gravitational wave detectors},''
  \href{http://dx.doi.org/10.1088/1361-6382/ad5488}{{\em Class. Quant. Grav.}
  {\bfseries 41} no.~14, (2024) 143001},
  \href{http://arxiv.org/abs/2401.07615}{{\ttfamily arXiv:2401.07615 [gr-qc]}}.

\bibitem{Chen:2024pge}
C.~Chen, A.~Ghoshal, G.~Tasinato, and E.~Tomberg, ``{Stochastic axionlike
  curvaton: Non-Gaussianity and primordial black holes without a large power
  spectrum},'' \href{http://dx.doi.org/10.1103/PhysRevD.111.063539}{{\em Phys.
  Rev. D} {\bfseries 111} no.~6, (2025) 063539},
  \href{http://arxiv.org/abs/2409.12950}{{\ttfamily arXiv:2409.12950
  [astro-ph.CO]}}.

\bibitem{Iovino:2025cdy}
A.~Iovino, Junior., G.~Perna, and H.~Veerm{\"a}e, ``{The impact of
  non-Gaussianity when searching for Primordial Black Holes with LISA},''
  \href{http://arxiv.org/abs/2512.13648}{{\ttfamily arXiv:2512.13648
  [astro-ph.CO]}}.

\bibitem{Zhong:2025xwm}
J.~Zhong, C.~Chen, and Y.-F. Cai, ``{Can asteroid-mass PBHDM be compatible with
  catalyzed phase transition interpretation of PTA?},''
  \href{http://dx.doi.org/10.1088/1475-7516/2025/10/033}{{\em JCAP} {\bfseries
  10} (2025) 033}, \href{http://arxiv.org/abs/2504.12105}{{\ttfamily
  arXiv:2504.12105 [astro-ph.CO]}}.

\bibitem{Ning:2026nfs}
Z.~Ning, X.-X. Zeng, R.-G. Cai, and S.-J. Wang, ``{Numerical simulations of
  primordial black hole formation via delayed first-order phase transitions},''
  \href{http://arxiv.org/abs/2601.21878}{{\ttfamily arXiv:2601.21878 [gr-qc]}}.

\bibitem{Ye:2026ffe}
X.~Ye, L.~F. Demetrio, E.~J. Barroso, S.-F. Yan, and N.~Pinto-Neto,
  ``{Primordial Black Hole Formation in Dust-Radiation Bouncing Cosmologies},''
  \href{http://arxiv.org/abs/2602.12057}{{\ttfamily arXiv:2602.12057 [gr-qc]}}.

\bibitem{Gangopadhyay:2023qjr}
M.~R. Gangopadhyay, V.~V. Godithi, R.~Inui, K.~Ichiki, T.~Kajino,
  A.~Manusankar, G.~J. Mathews, and Yogesh, ``{Is the NANOGrav detection
  evidence of resonant particle creation during inflation?},''
  \href{http://dx.doi.org/10.1016/j.jheap.2025.100358}{{\em JHEAp} {\bfseries
  47} (2025) 100358}, \href{http://arxiv.org/abs/2309.03101}{{\ttfamily
  arXiv:2309.03101 [astro-ph.CO]}}.

\bibitem{Yogesh:2025hll}
Yogesh and A.~Mohammadi, ``{Nonstandard Thermal History and Formation of
  Primordial Black Holes and SIGWs in
  Einstein{\textendash}Gauss{\textendash}Bonnet Gravity},''
  \href{http://dx.doi.org/10.3847/1538-4357/adcee5}{{\em Astrophys. J.}
  {\bfseries 986} no.~1, (2025) 35},
  \href{http://arxiv.org/abs/2501.01867}{{\ttfamily arXiv:2501.01867 [gr-qc]}}.

\bibitem{Mohammadi:2025avz}
A.~Mohammadi, Yogesh, Q.~Wu, and T.~Zhu, ``{Spinning Primordial Black Holes and
  Scalar Induced Gravitational Waves from Single Field Inflation},''
  \href{http://arxiv.org/abs/2512.05435}{{\ttfamily arXiv:2512.05435
  [astro-ph.CO]}}.

\bibitem{Allegrini:2024ooy}
S.~Allegrini, L.~Del~Grosso, A.~J. Iovino, and A.~Urbano, ``{Is the formation
  of primordial black holes from single-field inflation compatible with
  standard cosmology?},'' \href{http://dx.doi.org/10.1103/9rr3-ltbt}{{\em Phys.
  Rev. D} {\bfseries 111} no.~12, (2025) 123557},
  \href{http://arxiv.org/abs/2412.14049}{{\ttfamily arXiv:2412.14049
  [astro-ph.CO]}}.

\bibitem{Allegrini:2025jha}
S.~Allegrini, A.~J. Iovino, and H.~Veerm{\"a}e, ``{Beware of the running $n_s$
  when producing heavy primordial black holes},''
  \href{http://arxiv.org/abs/2510.18791}{{\ttfamily arXiv:2510.18791
  [astro-ph.CO]}}.

\bibitem{Ashoorioon:2018uey}
A.~Ashoorioon, R.~Casadio, M.~Cicoli, G.~Geshnizjani, and H.~J. Kim,
  ``{Extended Effective Field Theory of Inflation},''
  \href{http://dx.doi.org/10.1007/JHEP02(2018)172}{{\em JHEP} {\bfseries 02}
  (2018) 172}, \href{http://arxiv.org/abs/1802.03040}{{\ttfamily
  arXiv:1802.03040 [hep-th]}}.

\bibitem{Ashoorioon:2019xqc}
A.~Ashoorioon, A.~Rostami, and J.~T. Firouzjaee, ``{EFT compatible PBHs:
  effective spawning of the seeds for primordial black holes during
  inflation},'' \href{http://dx.doi.org/10.1007/JHEP07(2021)087}{{\em JHEP}
  {\bfseries 07} (2021) 087}, \href{http://arxiv.org/abs/1912.13326}{{\ttfamily
  arXiv:1912.13326 [astro-ph.CO]}}.

\bibitem{Ashoorioon:2020hln}
A.~Ashoorioon, A.~Rostami, and J.~T. Firouzjaee, ``{Examining the end of
  inflation with primordial black holes mass distribution and gravitational
  waves},'' \href{http://dx.doi.org/10.1103/PhysRevD.103.123512}{{\em Phys.
  Rev. D} {\bfseries 103} (2021) 123512},
  \href{http://arxiv.org/abs/2012.02817}{{\ttfamily arXiv:2012.02817
  [astro-ph.CO]}}.

\bibitem{Ashoorioon:2022raz}
A.~Ashoorioon, K.~Rezazadeh, and A.~Rostami, ``{NANOGrav signal from the end of
  inflation and the LIGO mass and heavier primordial black holes},''
  \href{http://dx.doi.org/10.1016/j.physletb.2022.137542}{{\em Phys. Lett. B}
  {\bfseries 835} (2022) 137542},
  \href{http://arxiv.org/abs/2202.01131}{{\ttfamily arXiv:2202.01131
  [astro-ph.CO]}}.

\bibitem{Cai:2023uhc}
Y.~Cai, M.~Zhu, and Y.-S. Piao, ``{Primordial Black Holes from Null Energy
  Condition Violation during Inflation},''
  \href{http://dx.doi.org/10.1103/PhysRevLett.133.021001}{{\em Phys. Rev.
  Lett.} {\bfseries 133} no.~2, (2024) 021001},
  \href{http://arxiv.org/abs/2305.10933}{{\ttfamily arXiv:2305.10933 [gr-qc]}}.

\bibitem{Cai:2020qpu}
Y.~Cai and Y.-S. Piao, ``{Intermittent null energy condition violations during
  inflation and primordial gravitational waves},''
  \href{http://dx.doi.org/10.1103/PhysRevD.103.083521}{{\em Phys. Rev. D}
  {\bfseries 103} no.~8, (2021) 083521},
  \href{http://arxiv.org/abs/2012.11304}{{\ttfamily arXiv:2012.11304 [gr-qc]}}.

\bibitem{Cai:2022nqv}
Y.~Cai and Y.-S. Piao, ``{Generating enhanced primordial GWs during inflation
  with intermittent violation of NEC and diminishment of GW propagating
  speed},'' \href{http://dx.doi.org/10.1007/JHEP06(2022)067}{{\em JHEP}
  {\bfseries 06} (2022) 067}, \href{http://arxiv.org/abs/2201.04552}{{\ttfamily
  arXiv:2201.04552 [gr-qc]}}.

\bibitem{Cai:2022lec}
Y.~Cai, ``{Generating enhanced parity-violating gravitational waves during
  inflation with violation of the null energy condition},''
  \href{http://dx.doi.org/10.1103/PhysRevD.107.063512}{{\em Phys. Rev. D}
  {\bfseries 107} no.~6, (2023) 063512},
  \href{http://arxiv.org/abs/2212.10893}{{\ttfamily arXiv:2212.10893 [gr-qc]}}.

\bibitem{Jiang:2023gfe}
J.-Q. Jiang, Y.~Cai, G.~Ye, and Y.-S. Piao, ``{Broken blue-tilted inflationary
  gravitational waves: a joint analysis of NANOGrav 15-year and BICEP/Keck 2018
  data},'' \href{http://dx.doi.org/10.1088/1475-7516/2024/05/004}{{\em JCAP}
  {\bfseries 05} (2024) 004}, \href{http://arxiv.org/abs/2307.15547}{{\ttfamily
  arXiv:2307.15547 [astro-ph.CO]}}.

\bibitem{Ye:2023tpz}
G.~Ye, M.~Zhu, and Y.~Cai, ``{Null energy condition violation during inflation
  and pulsar timing array observations},''
  \href{http://dx.doi.org/10.1007/JHEP02(2024)008}{{\em JHEP} {\bfseries 02}
  (2024) 008}, \href{http://arxiv.org/abs/2312.10685}{{\ttfamily
  arXiv:2312.10685 [gr-qc]}}.

\bibitem{Jiang:2024woi}
Z.-W. Jiang, Y.~Cai, F.~Wang, and Y.-S. Piao, ``{Parity-violating primordial
  gravitational waves from null energy condition violation},''
  \href{http://dx.doi.org/10.1007/JHEP09(2024)067}{{\em JHEP} {\bfseries 09}
  (2024) 067}, \href{http://arxiv.org/abs/2406.16549}{{\ttfamily
  arXiv:2406.16549 [astro-ph.CO]}}.

\bibitem{Pan:2024ydt}
S.~Pan, Y.~Cai, and Y.-S. Piao, ``{Climbing over the potential barrier during
  inflation via null energy condition violation},''
  \href{http://dx.doi.org/10.1140/epjc/s10052-024-13340-1}{{\em Eur. Phys. J.
  C} {\bfseries 84} no.~9, (2024) 976},
  \href{http://arxiv.org/abs/2404.12655}{{\ttfamily arXiv:2404.12655
  [astro-ph.CO]}}.

\bibitem{Chen:2024mwg}
Z.-C. Chen and L.~Liu, ``{Constraints on inflation with null energy condition
  violation from advanced LIGO and advanced Virgo's first three observing
  runs},'' \href{http://dx.doi.org/10.1088/1475-7516/2024/06/028}{{\em JCAP}
  {\bfseries 06} (2024) 028}, \href{http://arxiv.org/abs/2404.07075}{{\ttfamily
  arXiv:2404.07075 [gr-qc]}}.

\bibitem{Chen:2024jca}
Z.-C. Chen and L.~Liu, ``{Detecting a gravitational wave background from
  inflation with null energy condition violation: prospects for Taiji},''
  \href{http://dx.doi.org/10.1140/epjc/s10052-024-13529-4}{{\em Eur. Phys. J.
  C} {\bfseries 84} no.~11, (2024) 1176},
  \href{http://arxiv.org/abs/2404.08375}{{\ttfamily arXiv:2404.08375 [gr-qc]}}.

\bibitem{DOnofrio:2025bol}
S.~D'Onofrio, S.~Odintsov, and T.~Paul, ``{Fitting NANOGrav 15-year data and
  ACT data with modified inflation in entropic cosmology},''
  \href{http://arxiv.org/abs/2510.20484}{{\ttfamily arXiv:2510.20484 [gr-qc]}}.

\bibitem{DeLuca:2025uov}
V.~De~Luca, A.~J. Iovino, and A.~Riotto, ``{Primordial Black Hole Ringdown: the
  Irreducible Stochastic Gravitational Wave Background},''
  \href{http://arxiv.org/abs/2507.04083}{{\ttfamily arXiv:2507.04083 [gr-qc]}}.

\bibitem{Yuan:2025bdp}
C.~Yuan, Z.~Zhong, and Q.-G. Huang, ``{Whispers from the Early Universe: The
  Ringdown of Primordial Black Holes},''
  \href{http://arxiv.org/abs/2507.07665}{{\ttfamily arXiv:2507.07665
  [astro-ph.CO]}}.

\bibitem{Phinney:2001di}
E.~S. Phinney, ``{A Practical theorem on gravitational wave backgrounds},''
  \href{http://arxiv.org/abs/astro-ph/0108028}{{\ttfamily
  arXiv:astro-ph/0108028}}.

\bibitem{Berti:2005ys}
E.~Berti, V.~Cardoso, and C.~M. Will, ``{On gravitational-wave spectroscopy of
  massive black holes with the space interferometer LISA},''
  \href{http://dx.doi.org/10.1103/PhysRevD.73.064030}{{\em Phys. Rev. D}
  {\bfseries 73} (2006) 064030},
  \href{http://arxiv.org/abs/gr-qc/0512160}{{\ttfamily arXiv:gr-qc/0512160}}.

\bibitem{Bruegmann:2006ulg}
B.~Bruegmann, J.~A. Gonzalez, M.~Hannam, S.~Husa, U.~Sperhake, and W.~Tichy,
  ``{Calibration of Moving Puncture Simulations},''
  \href{http://dx.doi.org/10.1103/PhysRevD.77.024027}{{\em Phys. Rev. D}
  {\bfseries 77} (2008) 024027},
  \href{http://arxiv.org/abs/gr-qc/0610128}{{\ttfamily arXiv:gr-qc/0610128}}.

\bibitem{Ajith:2007kx}
P.~Ajith {\em et~al.}, ``{A Template bank for gravitational waveforms from
  coalescing binary black holes. I. Non-spinning binaries},''
  \href{http://dx.doi.org/10.1103/PhysRevD.77.104017}{{\em Phys. Rev. D}
  {\bfseries 77} (2008) 104017},
  \href{http://arxiv.org/abs/0710.2335}{{\ttfamily arXiv:0710.2335 [gr-qc]}}.
  [Erratum: Phys.Rev.D 79, 129901 (2009)].

\bibitem{Reisswig:2010di}
C.~Reisswig and D.~Pollney, ``{Notes on the integration of numerical relativity
  waveforms},'' \href{http://dx.doi.org/10.1088/0264-9381/28/19/195015}{{\em
  Class. Quant. Grav.} {\bfseries 28} (2011) 195015},
  \href{http://arxiv.org/abs/1006.1632}{{\ttfamily arXiv:1006.1632 [gr-qc]}}.

\bibitem{Zhu:2011bd}
X.-J. Zhu, E.~Howell, T.~Regimbau, D.~Blair, and Z.-H. Zhu, ``{Stochastic
  Gravitational Wave Background from Coalescing Binary Black Holes},''
  \href{http://dx.doi.org/10.1088/0004-637X/739/2/86}{{\em Astrophys. J.}
  {\bfseries 739} (2011) 86}, \href{http://arxiv.org/abs/1104.3565}{{\ttfamily
  arXiv:1104.3565 [gr-qc]}}.

\bibitem{Zhu:2012xw}
X.-J. Zhu, E.~J. Howell, D.~G. Blair, and Z.-H. Zhu, ``{On the gravitational
  wave background from compact binary coalescences in the band of ground-based
  interferometers},'' \href{http://dx.doi.org/10.1093/mnras/stt207}{{\em Mon.
  Not. Roy. Astron. Soc.} {\bfseries 431} no.~1, (2013) 882--899},
  \href{http://arxiv.org/abs/1209.0595}{{\ttfamily arXiv:1209.0595 [gr-qc]}}.

\bibitem{Berti:2014bla}
E.~Berti, ``{A Black-Hole Primer: Particles, Waves, Critical Phenomena and
  Superradiant Instabilities},''
\newblock 10, 2014.
\newblock \href{http://arxiv.org/abs/1410.4481}{{\ttfamily arXiv:1410.4481
  [gr-qc]}}.

\bibitem{Raidal:2018bbj}
M.~Raidal, C.~Spethmann, V.~Vaskonen, and H.~Veerm{\"a}e, ``{Formation and
  Evolution of Primordial Black Hole Binaries in the Early Universe},''
  \href{http://dx.doi.org/10.1088/1475-7516/2019/02/018}{{\em JCAP} {\bfseries
  02} (2019) 018}, \href{http://arxiv.org/abs/1812.01930}{{\ttfamily
  arXiv:1812.01930 [astro-ph.CO]}}.

\bibitem{Vaskonen:2019jpv}
V.~Vaskonen and H.~Veerm{\"a}e, ``{Lower bound on the primordial black hole
  merger rate},'' \href{http://dx.doi.org/10.1103/PhysRevD.101.043015}{{\em
  Phys. Rev. D} {\bfseries 101} no.~4, (2020) 043015},
  \href{http://arxiv.org/abs/1908.09752}{{\ttfamily arXiv:1908.09752
  [astro-ph.CO]}}.

\bibitem{Cai:2019cdl}
R.-G. Cai, S.~Pi, and M.~Sasaki, ``{Universal infrared scaling of gravitational
  wave background spectra},''
  \href{http://dx.doi.org/10.1103/PhysRevD.102.083528}{{\em Phys. Rev. D}
  {\bfseries 102} no.~8, (2020) 083528},
  \href{http://arxiv.org/abs/1909.13728}{{\ttfamily arXiv:1909.13728
  [astro-ph.CO]}}.

\bibitem{Hutsi:2020sol}
G.~H{\"u}tsi, M.~Raidal, V.~Vaskonen, and H.~Veerm{\"a}e, ``{Two populations of
  LIGO-Virgo black holes},''
  \href{http://dx.doi.org/10.1088/1475-7516/2021/03/068}{{\em JCAP} {\bfseries
  03} (2021) 068}, \href{http://arxiv.org/abs/2012.02786}{{\ttfamily
  arXiv:2012.02786 [astro-ph.CO]}}.

\bibitem{Franciolini:2022tfm}
G.~Franciolini, I.~Musco, P.~Pani, and A.~Urbano, ``{From inflation to black
  hole mergers and back again: Gravitational-wave data-driven constraints on
  inflationary scenarios with a first-principle model of primordial black holes
  across the QCD epoch},''
  \href{http://dx.doi.org/10.1103/PhysRevD.106.123526}{{\em Phys. Rev. D}
  {\bfseries 106} no.~12, (2022) 123526},
  \href{http://arxiv.org/abs/2209.05959}{{\ttfamily arXiv:2209.05959
  [astro-ph.CO]}}.

\bibitem{Zhong:2024ysg}
Z.~Zhong, V.~Cardoso, and Y.~Chen, ``{Dynamical Lensing Tomography of Black
  Hole Ringdowns},''
  \href{http://dx.doi.org/10.1103/PhysRevLett.134.211402}{{\em Phys. Rev.
  Lett.} {\bfseries 134} no.~21, (2025) 211402},
  \href{http://arxiv.org/abs/2408.10303}{{\ttfamily arXiv:2408.10303 [gr-qc]}}.

\bibitem{Raidal:2024bmm}
M.~Raidal, V.~Vaskonen, and H.~Veerm{\"a}e, {\em {Formation of~Primordial Black
  Hole Binaries and~Their Merger Rates}}.
\newblock 2025.
\newblock \href{http://arxiv.org/abs/2404.08416}{{\ttfamily arXiv:2404.08416
  [astro-ph.CO]}}.

\bibitem{Hu:2026sxj}
H.-W. Hu, C.-J. Fang, and Z.-K. Guo, ``{Waveform stability of black hole
  ringdown with stochastic horizon structure},''
  \href{http://arxiv.org/abs/2602.08034}{{\ttfamily arXiv:2602.08034 [gr-qc]}}.

\bibitem{Berti:2009kk}
E.~Berti, V.~Cardoso, and A.~O. Starinets, ``{Quasinormal modes of black holes
  and black branes},''
  \href{http://dx.doi.org/10.1088/0264-9381/26/16/163001}{{\em Class. Quant.
  Grav.} {\bfseries 26} (2009) 163001},
  \href{http://arxiv.org/abs/0905.2975}{{\ttfamily arXiv:0905.2975 [gr-qc]}}.

\bibitem{Berti:2025hly}
J.~Abedi {\em et~al.}, ``{Black hole spectroscopy: from theory to
  experiment},'' \href{http://arxiv.org/abs/2505.23895}{{\ttfamily
  arXiv:2505.23895 [gr-qc]}}.

\bibitem{Rubakov:2014jja}
V.~A. Rubakov, ``{The Null Energy Condition and its violation},''
  \href{http://dx.doi.org/10.3367/UFNe.0184.201402b.0137}{{\em Phys. Usp.}
  {\bfseries 57} (2014) 128--142},
  \href{http://arxiv.org/abs/1401.4024}{{\ttfamily arXiv:1401.4024 [hep-th]}}.

\bibitem{Curiel:2014zba}
E.~Curiel, ``{A Primer on Energy Conditions},''
  \href{http://dx.doi.org/10.1007/978-1-4939-3210-8_3}{{\em Einstein Stud.}
  {\bfseries 13} (2017) 43--104},
  \href{http://arxiv.org/abs/1405.0403}{{\ttfamily arXiv:1405.0403
  [physics.hist-ph]}}.

\bibitem{Kontou:2020bta}
E.-A. Kontou and K.~Sanders, ``{Energy conditions in general relativity and
  quantum field theory},''
  \href{http://dx.doi.org/10.1088/1361-6382/ab8fcf}{{\em Class. Quant. Grav.}
  {\bfseries 37} no.~19, (2020) 193001},
  \href{http://arxiv.org/abs/2003.01815}{{\ttfamily arXiv:2003.01815 [gr-qc]}}.

\bibitem{Cai:2016thi}
Y.~Cai, Y.~Wan, H.-G. Li, T.~Qiu, and Y.-S. Piao, ``{The Effective Field Theory
  of nonsingular cosmology},''
  \href{http://dx.doi.org/10.1007/JHEP01(2017)090}{{\em JHEP} {\bfseries 01}
  (2017) 090}, \href{http://arxiv.org/abs/1610.03400}{{\ttfamily
  arXiv:1610.03400 [gr-qc]}}.

\bibitem{Creminelli:2016zwa}
P.~Creminelli, D.~Pirtskhalava, L.~Santoni, and E.~Trincherini, ``{Stability of
  Geodesically Complete Cosmologies},''
  \href{http://dx.doi.org/10.1088/1475-7516/2016/11/047}{{\em JCAP} {\bfseries
  11} (2016) 047}, \href{http://arxiv.org/abs/1610.04207}{{\ttfamily
  arXiv:1610.04207 [hep-th]}}.

\bibitem{Cai:2017tku}
Y.~Cai, H.-G. Li, T.~Qiu, and Y.-S. Piao, ``{The Effective Field Theory of
  nonsingular cosmology: II},''
  \href{http://dx.doi.org/10.1140/epjc/s10052-017-4938-y}{{\em Eur. Phys. J. C}
  {\bfseries 77} no.~6, (2017) 369},
  \href{http://arxiv.org/abs/1701.04330}{{\ttfamily arXiv:1701.04330 [gr-qc]}}.

\bibitem{Cai:2017dyi}
Y.~Cai and Y.-S. Piao, ``{A covariant Lagrangian for stable nonsingular
  bounce},'' \href{http://dx.doi.org/10.1007/JHEP09(2017)027}{{\em JHEP}
  {\bfseries 09} (2017) 027}, \href{http://arxiv.org/abs/1705.03401}{{\ttfamily
  arXiv:1705.03401 [gr-qc]}}.

\bibitem{Kolevatov:2017voe}
R.~Kolevatov, S.~Mironov, N.~Sukhov, and V.~Volkova, ``{Cosmological bounce and
  Genesis beyond Horndeski},''
  \href{http://dx.doi.org/10.1088/1475-7516/2017/08/038}{{\em JCAP} {\bfseries
  08} (2017) 038}, \href{http://arxiv.org/abs/1705.06626}{{\ttfamily
  arXiv:1705.06626 [hep-th]}}.

\bibitem{Ilyas:2020qja}
A.~Ilyas, M.~Zhu, Y.~Zheng, Y.-F. Cai, and E.~N. Saridakis, ``{DHOST Bounce},''
  \href{http://dx.doi.org/10.1088/1475-7516/2020/09/002}{{\em JCAP} {\bfseries
  09} (2020) 002}, \href{http://arxiv.org/abs/2002.08269}{{\ttfamily
  arXiv:2002.08269 [gr-qc]}}.

\bibitem{Ijjas:2016tpn}
A.~Ijjas and P.~J. Steinhardt, ``{Classically stable nonsingular cosmological
  bounces},'' \href{http://dx.doi.org/10.1103/PhysRevLett.117.121304}{{\em
  Phys. Rev. Lett.} {\bfseries 117} no.~12, (2016) 121304},
  \href{http://arxiv.org/abs/1606.08880}{{\ttfamily arXiv:1606.08880 [gr-qc]}}.

\bibitem{Ijjas:2016vtq}
A.~Ijjas and P.~J. Steinhardt, ``{Fully stable cosmological solutions with a
  non-singular classical bounce},''
  \href{http://dx.doi.org/10.1016/j.physletb.2016.11.047}{{\em Phys. Lett. B}
  {\bfseries 764} (2017) 289--294},
  \href{http://arxiv.org/abs/1609.01253}{{\ttfamily arXiv:1609.01253 [gr-qc]}}.

\bibitem{Cai:2017dxl}
Y.~Cai and Y.-S. Piao, ``{Higher order derivative coupling to gravity and its
  cosmological implications},''
  \href{http://dx.doi.org/10.1103/PhysRevD.96.124028}{{\em Phys. Rev. D}
  {\bfseries 96} no.~12, (2017) 124028},
  \href{http://arxiv.org/abs/1707.01017}{{\ttfamily arXiv:1707.01017 [gr-qc]}}.

\bibitem{Cai:2017pga}
Y.~Cai, Y.-T. Wang, J.-Y. Zhao, and Y.-S. Piao, ``{Primordial perturbations
  with pre-inflationary bounce},''
  \href{http://dx.doi.org/10.1103/PhysRevD.97.103535}{{\em Phys. Rev. D}
  {\bfseries 97} no.~10, (2018) 103535},
  \href{http://arxiv.org/abs/1709.07464}{{\ttfamily arXiv:1709.07464
  [astro-ph.CO]}}.

\bibitem{Cai:2022ori}
Y.~Cai, J.~Xu, S.~Zhao, and S.~Zhou, ``{Perturbative unitarity and NEC
  violation in genesis cosmology},''
  \href{http://dx.doi.org/10.1007/JHEP10(2022)140}{{\em JHEP} {\bfseries 10}
  (2022) 140}, \href{http://arxiv.org/abs/2207.11772}{{\ttfamily
  arXiv:2207.11772 [gr-qc]}}. [Erratum: JHEP 11, 063 (2022)].

\bibitem{Yu:2025wak}
D.-H. Yu, M.~Zhu, and Y.~Cai, ``{Constraints on Genesis Cosmology from the
  Smeared Null Energy Condition},''
  \href{http://arxiv.org/abs/2512.04934}{{\ttfamily arXiv:2512.04934 [gr-qc]}}.

\bibitem{Mironov:2018oec}
S.~Mironov, V.~Rubakov, and V.~Volkova, ``{Bounce beyond Horndeski with GR
  asymptotics and $\gamma$-crossing},''
  \href{http://dx.doi.org/10.1088/1475-7516/2018/10/050}{{\em JCAP} {\bfseries
  10} (2018) 050}, \href{http://arxiv.org/abs/1807.08361}{{\ttfamily
  arXiv:1807.08361 [hep-th]}}.

\bibitem{Ye:2019sth}
G.~Ye and Y.-S. Piao, ``{Bounce in general relativity and higher-order
  derivative operators},''
  \href{http://dx.doi.org/10.1103/PhysRevD.99.084019}{{\em Phys. Rev. D}
  {\bfseries 99} no.~8, (2019) 084019},
  \href{http://arxiv.org/abs/1901.08283}{{\ttfamily arXiv:1901.08283 [gr-qc]}}.

\bibitem{Nandi:2019xag}
D.~Nandi and L.~Sriramkumar, ``{Can a nonminimal coupling restore the
  consistency condition in bouncing universes?},''
  \href{http://dx.doi.org/10.1103/PhysRevD.101.043506}{{\em Phys. Rev. D}
  {\bfseries 101} no.~4, (2020) 043506},
  \href{http://arxiv.org/abs/1904.13254}{{\ttfamily arXiv:1904.13254 [gr-qc]}}.

\bibitem{Nandi:2020szp}
D.~Nandi, ``{Stability of a viable non-minimal bounce},''
  \href{http://dx.doi.org/10.3390/universe7030062}{{\em Universe} {\bfseries 7}
  no.~3, (2021) 62}, \href{http://arxiv.org/abs/2009.03134}{{\ttfamily
  arXiv:2009.03134 [gr-qc]}}.

\bibitem{Nandi:2023ooo}
D.~Nandi and M.~Kaur, ``{Inflation vs. Ekpyrosis {\textemdash} Comparing
  stability in general non-minimal theory},''
  \href{http://dx.doi.org/10.1016/j.dark.2024.101430}{{\em Phys. Dark Univ.}
  {\bfseries 44} (2024) 101430},
  \href{http://arxiv.org/abs/2302.03413}{{\ttfamily arXiv:2302.03413
  [astro-ph.CO]}}.

\bibitem{Zhu:2021whu}
M.~Zhu, A.~Ilyas, Y.~Zheng, Y.-F. Cai, and E.~N. Saridakis, ``{Scalar and
  tensor perturbations in DHOST bounce cosmology},''
  \href{http://dx.doi.org/10.1088/1475-7516/2021/11/045}{{\em JCAP} {\bfseries
  11} no.~11, (2021) 045}, \href{http://arxiv.org/abs/2108.01339}{{\ttfamily
  arXiv:2108.01339 [gr-qc]}}.

\bibitem{Zhu:2021ggm}
M.~Zhu and Y.~Zheng, ``{Improved DHOST Genesis},''
  \href{http://dx.doi.org/10.1007/JHEP11(2021)163}{{\em JHEP} {\bfseries 11}
  (2021) 163}, \href{http://arxiv.org/abs/2109.05277}{{\ttfamily
  arXiv:2109.05277 [gr-qc]}}.

\bibitem{Qiu:2024sdd}
T.~Qiu and M.~Zhu, ``{Interpreting pulsar timing array data of gravitational
  waves with Ekpyrosis-bouncing cosmology},''
  \href{http://dx.doi.org/10.1103/PhysRevD.111.043508}{{\em Phys. Rev. D}
  {\bfseries 111} no.~4, (2025) 043508},
  \href{http://arxiv.org/abs/2408.06582}{{\ttfamily arXiv:2408.06582 [gr-qc]}}.

\bibitem{Ageeva:2024knc}
Y.~Ageeva, M.~Kotenko, and P.~Petrov, ``{Primordial non-Gaussianities for
  nonsingular Horndeski cosmologies},''
  \href{http://dx.doi.org/10.1103/ghz6-w27b}{{\em Phys. Rev. D} {\bfseries 112}
  no.~6, (2025) 063526}, \href{http://arxiv.org/abs/2410.10742}{{\ttfamily
  arXiv:2410.10742 [hep-th]}}.

\bibitem{Akama:2025ows}
S.~Akama, ``{Primordial full bispectra from the general bounce cosmology},''
  \href{http://dx.doi.org/10.1088/1475-7516/2025/06/063}{{\em JCAP} {\bfseries
  06} (2025) 063}, \href{http://arxiv.org/abs/2502.14850}{{\ttfamily
  arXiv:2502.14850 [astro-ph.CO]}}.

\bibitem{Dehghani:2025udv}
A.~Dehghani, G.~Geshnizjani, and J.~Quintin, ``{Cuscuton bounce beyond the
  linear regime: bispectrum and strong coupling constraints},''
  \href{http://dx.doi.org/10.1088/1475-7516/2025/05/026}{{\em JCAP} {\bfseries
  05} (2025) 026}, \href{http://arxiv.org/abs/2503.01992}{{\ttfamily
  arXiv:2503.01992 [hep-th]}}.

\bibitem{Cai:2016ihp}
Y.~Cai, Y.-T. Wang, and Y.-S. Piao, ``{Chirality oscillation of primordial
  gravitational waves during inflation},''
  \href{http://dx.doi.org/10.1007/JHEP03(2017)024}{{\em JHEP} {\bfseries 03}
  (2017) 024}, \href{http://arxiv.org/abs/1608.06508}{{\ttfamily
  arXiv:1608.06508 [astro-ph.CO]}}.

\bibitem{Zhu:2023lhv}
M.~Zhu and Y.~Cai, ``{Parity-violation in bouncing cosmology},''
  \href{http://dx.doi.org/10.1007/JHEP04(2023)095}{{\em JHEP} {\bfseries 04}
  (2023) 095}, \href{http://arxiv.org/abs/2301.13502}{{\ttfamily
  arXiv:2301.13502 [gr-qc]}}.

\bibitem{Goldberg:1966uu}
J.~N. Goldberg, A.~J. MacFarlane, E.~T. Newman, F.~Rohrlich, and E.~C.~G.
  Sudarshan, ``{Spin-$s$ spherical harmonics and $\eth$},''
  \href{http://dx.doi.org/10.1063/1.1705135}{{\em J. Math. Phys.} {\bfseries 8}
  (1967) 2155}.

\bibitem{Planck:2018vyg}
{\bfseries Planck} Collaboration, N.~Aghanim {\em et~al.}, ``{Planck 2018
  results. VI. Cosmological parameters},''
  \href{http://dx.doi.org/10.1051/0004-6361/201833910}{{\em Astron. Astrophys.}
  {\bfseries 641} (2020) A6}, \href{http://arxiv.org/abs/1807.06209}{{\ttfamily
  arXiv:1807.06209 [astro-ph.CO]}}. [Erratum: Astron.Astrophys. 652, C4
  (2021)].

\bibitem{Roshan:2024qnv}
R.~Roshan and G.~White, ``{Using gravitational waves to see the first second of
  the Universe},'' \href{http://dx.doi.org/10.1103/RevModPhys.97.015001}{{\em
  Rev. Mod. Phys.} {\bfseries 97} no.~1, (2025) 015001},
  \href{http://arxiv.org/abs/2401.04388}{{\ttfamily arXiv:2401.04388
  [hep-ph]}}.

\bibitem{Cooke:2013cba}
R.~Cooke, M.~Pettini, R.~A. Jorgenson, M.~T. Murphy, and C.~C. Steidel,
  ``{Precision measures of the primordial abundance of deuterium},''
  \href{http://dx.doi.org/10.1088/0004-637X/781/1/31}{{\em Astrophys. J.}
  {\bfseries 781} no.~1, (2014) 31},
  \href{http://arxiv.org/abs/1308.3240}{{\ttfamily arXiv:1308.3240
  [astro-ph.CO]}}.

\bibitem{Clarke:2020bil}
T.~J. Clarke, E.~J. Copeland, and A.~Moss, ``{Constraints on primordial
  gravitational waves from the Cosmic Microwave Background},''
  \href{http://dx.doi.org/10.1088/1475-7516/2020/10/002}{{\em JCAP} {\bfseries
  10} (2020) 002}, \href{http://arxiv.org/abs/2004.11396}{{\ttfamily
  arXiv:2004.11396 [astro-ph.CO]}}.

\bibitem{data260220}
\url{https://github.com/JZZhang05/Data-PBH-SGWB}.

\end{thebibliography}\endgroup
\bibliographystyle{utphys}

\end{document}